\documentstyle[twoside]{article}

\catcode`\@=11
\long\def\@makefntext#1{
\protect\noindent \hbox to 3.2pt {\hskip-.9pt  
$^{{\eightrm\@thefnmark}}$\hfil}#1\hfill}		

\def\thefootnote{\fnsymbol{footnote}}
\def\@makefnmark{\hbox to 0pt{$^{\@thefnmark}$\hss}}	
	
\def\ps@myheadings{\let\@mkboth\@gobbletwo
\def\@oddhead{\hbox{}
\rightmark\hfil\eightrm\thepage}   
\def\@oddfoot{}\def\@evenhead{\eightrm\thepage\hfil
\leftmark\hbox{}}\def\@evenfoot{}
\def\sectionmark##1{}\def\subsectionmark##1{}}

\oddsidemargin=\evensidemargin
\renewcommand{\thefootnote}{\fnsymbol{footnote}}

\newcounter{sectionc}\newcounter{subsectionc}\newcounter{subsubsectionc}
\renewcommand{\section}[1] {\vspace{12pt}\addtocounter{sectionc}{1} 
\setcounter{subsectionc}{0}\setcounter{subsubsectionc}{0}\noindent 
	{\tenbf\thesectionc. #1}\par\vspace{5pt}}
\renewcommand{\subsection}[1] {\vspace{12pt}\addtocounter{subsectionc}{1} 
	\setcounter{subsubsectionc}{0}\noindent 
	{\bf\thesectionc.\thesubsectionc. {\kern1pt \bfit #1}}\par\vspace{5pt}}
\renewcommand{\subsubsection}[1] {\vspace{12pt}\addtocounter{subsubsectionc}{1}
	\noindent{\tenrm\thesectionc.\thesubsectionc.\thesubsubsectionc.
	{\kern1pt \tenit #1}}\par\vspace{5pt}}
\newcommand{\nonumsection}[1] {\vspace{12pt}\noindent{\tenbf #1}
	\par\vspace{5pt}}

\newcounter{appendixc}
\newcounter{subappendixc}[appendixc]
\newcounter{subsubappendixc}[subappendixc]
\renewcommand{\thesubappendixc}{\Alph{appendixc}.\arabic{subappendixc}}
\renewcommand{\thesubsubappendixc}
	{\Alph{appendixc}.\arabic{subappendixc}.\arabic{subsubappendixc}}

\renewcommand{\appendix}[1] {\vspace{12pt}
        \refstepcounter{appendixc}
        \setcounter{figure}{0}
        \setcounter{table}{0}
        \setcounter{lemma}{0}
        \setcounter{theorem}{0}
        \setcounter{corollary}{0}
        \setcounter{definition}{0}
        \setcounter{equation}{0}
        \renewcommand{\thefigure}{\Alph{appendixc}.\arabic{figure}}
        \renewcommand{\thetable}{\Alph{appendixc}.\arabic{table}}
        \renewcommand{\theappendixc}{\Alph{appendixc}}
        \renewcommand{\thelemma}{\Alph{appendixc}.\arabic{lemma}}
        \renewcommand{\thetheorem}{\Alph{appendixc}.\arabic{theorem}}
        \renewcommand{\thedefinition}{\Alph{appendixc}.\arabic{definition}}
        \renewcommand{\thecorollary}{\Alph{appendixc}.\arabic{corollary}}
        \renewcommand{\theequation}{\Alph{appendixc}.\arabic{equation}}
        \noindent{\tenbf Appendix \theappendixc #1}\par\vspace{5pt}}
\newcommand{\subappendix}[1] {\vspace{12pt}
        \refstepcounter{subappendixc}
        \noindent{\bf Appendix \thesubappendixc. {\kern1pt \bfit #1}}
	\par\vspace{5pt}}
\newcommand{\subsubappendix}[1] {\vspace{12pt}
        \refstepcounter{subsubappendixc}
        \noindent{\rm Appendix \thesubsubappendixc. {\kern1pt \tenit #1}}
	\par\vspace{5pt}}

\newcommand{\textlineskip}{\baselineskip=13pt}
\newcommand{\smalllineskip}{\baselineskip=10pt}

\def\eightcirc{
\begin{picture}(0,0)
\put(4.4,1.8){\circle{6.5}}
\end{picture}}
\def\eightcopyright{\eightcirc\kern2.7pt\hbox{\eightrm c}}

\def\abstracts#1#2#3{{
	\centering{\begin{minipage}{4.5in}\footnotesize\baselineskip=10pt
	\parindent=0pt #1\par 
	\parindent=15pt #2\par
	\parindent=15pt #3
	\end{minipage}}\par}} 

\newcommand{\bibit}{\nineit}

\renewenvironment{thebibliography}[1]
	{\frenchspacing
	 \ninerm\baselineskip=11pt
	 \begin{list}{\arabic{enumi}.}
	{\usecounter{enumi}\setlength{\parsep}{0pt}
	 \setlength{\leftmargin 12.7pt}{\rightmargin 0pt} 
	 \setlength{\itemsep}{0pt} \settowidth
	{\labelwidth}{#1.}\sloppy}}{\end{list}}

\newcounter{itemlistc}
\newcounter{romanlistc}
\newcounter{alphlistc}
\newcounter{arabiclistc}

\newcommand{\fcaption}[1]{
        \refstepcounter{figure}
        \setbox\@tempboxa = \hbox{\footnotesize Fig.~\thefigure. #1}
        \ifdim \wd\@tempboxa > 5in
           {\begin{center}
        \parbox{5in}{\footnotesize\smalllineskip Fig.~\thefigure. #1}
            \end{center}}
        \else
             {\begin{center}
             {\footnotesize Fig.~\thefigure. #1}
              \end{center}}
        \fi}

\newcommand{\tcaption}[1]{
        \refstepcounter{table}
        \setbox\@tempboxa = \hbox{\footnotesize Table~\thetable. #1}
        \ifdim \wd\@tempboxa > 5in
           {\begin{center}
        \parbox{5in}{\footnotesize\smalllineskip Table~\thetable. #1}
            \end{center}}
        \else
             {\begin{center}
             {\footnotesize Table~\thetable. #1}
              \end{center}}
        \fi}

\def\@citex[#1]#2{\if@filesw\immediate\write\@auxout
	{\string\citation{#2}}\fi
\def\@citea{}\@cite{\@for\@citeb:=#2\do
	{\@citea\def\@citea{,}\@ifundefined
	{b@\@citeb}{{\bf ?}\@warning
	{Citation `\@citeb' on page \thepage \space undefined}}
	{\csname b@\@citeb\endcsname}}}{#1}}

\newif\if@cghi
\def\cite{\@cghitrue\@ifnextchar [{\@tempswatrue
	\@citex}{\@tempswafalse\@citex[]}}
\def\citelow{\@cghifalse\@ifnextchar [{\@tempswatrue
	\@citex}{\@tempswafalse\@citex[]}}
\def\@cite#1#2{{$\null^{#1}$\if@tempswa\typeout
	{IJCGA warning: optional citation argument 
	ignored: `#2'} \fi}}

\def\pmb#1{\setbox0=\hbox{#1}
	\kern-.025em\copy0\kern-\wd0
	\kern.05em\copy0\kern-\wd0
	\kern-.025em\raise.0433em\box0}

\def\fnt#1#2{\footnotetext{\kern-.3em
	{$^{\mbox{\scriptsize #1}}$}{#2}}}

\def\thefootnote{\fnsymbol{footnote}}
\def\@makefnmark{\hbox to 0pt{$^{\@thefnmark}$\hss}}	
	
\def\ps@myheadings{%
    \let\@oddfoot\@empty\let\@evenfoot\@empty
    \def\@evenhead{\slshape\leftmark\hfil}
    \def\@oddhead{\hfil{\slshape\rightmark}}
    \let\@mkboth\@gobbletwo
    \let\sectionmark\@gobble
    \let\subsectionmark\@gobble
    }
\font\tenrm=cmr10
\font\tenit=cmti10 
\font\tenbf=cmbx10
\font\bfit=cmbxti10 at 10pt
\font\ninerm=cmr9
\font\nineit=cmti9

\font\eightrm=cmr8

\def\qed{\hbox{${\vcenter{\vbox{                        
   \hrule height 0.4pt\hbox{\vrule width 0.4pt height 6pt
   \kern5pt\vrule width 0.4pt}\hrule height 0.4pt}}}$}}

\renewcommand{\thefootnote}{\fnsymbol{footnote}}  

\begin{document}
\setlength{\textheight}{7.7truein}  
\thispagestyle{empty}
\normalsize\textlineskip
\setcounter{page}{1}
\vspace*{0.88truein}

\centerline{\bf DYNAMICS OF A GENERALIZED}
\vspace*{0.035truein}
\centerline{\bf COSMOLOGICAL SCALAR-TENSOR THEORY}
\vspace*{0.37truein}

\centerline{\footnotesize TAKAO FUKUI,\footnote{E-mail:
   fukui@sciborg.uwaterloo.ca} \footnote{Permanent address: Department of
   Language and Culture, Dokkyo University, Soka, Saitama 340-0042, Japan}
   \hspace{5pt}
   JAMES M. OVERDUIN\footnote{Current address: Astrophysics and Cosmology
   Group, Department of Physics, Waseda University, Okubo 3-4-1,
   Shinjuku-ku, Tokyo 169-8555, Japan}}
\baselineskip=12pt
\centerline{\footnotesize\it Department of Physics, University of Waterloo}
\baselineskip=10pt
\centerline{\footnotesize\it Waterloo, Ontario, Canada N2L 3G1}
\vspace*{0.225truein}

\vspace*{0.21truein}
\abstracts{A generalized scalar-tensor (GST) theory is investigated
whose cosmological (or quintessence) term depends on both a scalar
field and its time derivative.  A correspondence with solutions of
five-dimensional Space-Time-Matter (STM) theory is noted.
Analytic solutions are found for the scale factor, scalar field and
cosmological term.  Models with free parameters of order unity are
consistent with recent observational data and could be relevant to
both the dark matter and cosmological ``constant'' problems.}{}{}

\vspace*{1pt}\textlineskip
\section{Introduction}
\vspace*{-0.5pt}
\noindent
Generalized scalar-tensor (GST) theories\cite{ref1} have enjoyed new
attention in recent years, partly because the potential energy of the
scalar field and/or the presence of a variable cosmological term could
drive inflation, resolving puzzles such as the monopole, horizon and
flatness problems.\cite{ref2}  The variable cosmological term has also
been mentioned as a possible solution to the cosmological ``constant''
problem,\cite{ref3} and --- most recently --- as a candidate for the
dark matter (or quintessence) making up most of the Universe.\cite{ref4}

In a previous paper\cite{ref5} we have introduced a modified GST theory
in which the cosmological term $\Lambda$ depends not only on the scalar
field $\phi$ but its gradient $\phi_{,i}\phi^{,i}$ as well.
A correspondence with solutions of Wesson's five-dimensional
Space-Time-Matter (STM) theory\cite{ref6} has been observed, with the
scalar field arising in four-dimensional spacetime as a manifestation
of higher-dimensional geometry.  Here we obtain cosmological solutions
of the theory and explore their properties in detail, with special
regard to constraints from recent observational data.

The theory is summarized and applied to cosmology (the Robertson-Walker
metric) in Section~2.  The connection to STM theory is drawn in Section~3,
where solutions for the scale factor $a$, Hubble parameter $H$ and
deceleration parameter $q$ are discussed.  Analytic expressions for the
scalar field and cosmological term are derived in Sections~4 and 5, and
related to observation via the energy density parameters $\Omega_{m,0}$
and $\Omega_{\Lambda,0}$.  Section~6 is a discussion.

\setcounter{footnote}{0}
\renewcommand{\thefootnote}{\alph{footnote}}

\section{Modified GST Cosmology}
\noindent
We consider here a modified GST theory, first introduced by the authors
in Ref.~5, in which $\Lambda$ depends on both the scalar field $\phi$
and its gradient $\phi_{,i}\phi^{,i}$ (the Brans-Dicke coupling parameter
$\omega$ is however taken as a function of $\phi$ alone).  This dependence
leads to significant departures from earlier GST theories in which
$\Lambda=\Lambda(\phi)$ only.\cite{ref7}  Possible implications for the
early universe have recently been investigated in Ref.~8.  In this paper
we focus primarily on the later stages of cosmic evolution (i.e., the
radiation and matter-dominated eras).

The action principle of the modified GST theory is
\begin{equation}
0 = \delta \int \left\{ \phi \left[ R + 2\Lambda (\phi, \phi_{,i}
   \phi^{,i}) \right] + \frac{16\pi}{c^4} L_m - \omega(\phi)
   \frac{\phi_{,i} \phi^{,i}}{\phi} \right\} \sqrt{-g} \, d\Omega \; ,
\label{GSTaction}
\end{equation}
where the Latin index $i$ runs over 0,1,2 and 3.  Variation leads to 
the following field equation for the metric field $g_{ij}$
\begin{equation}
R_{ij} - \frac{1}{2} g_{ij} R = \frac{8\pi}{\phi c^4} \left(
   T_{ij} + T^{\phi}_{ij} \right) \; ,
\label{FEgij}
\end{equation}
where $b\equiv\phi_{,l}\phi^{,l}$, $T_{ij}$ is the energy-momentum tensor
of matter, and the energy-momentum tensor of the scalar field is defined
as follows\cite{ref9}
\begin{equation}
\frac{8\pi}{\phi c^4} T^{\phi}_{ij} \equiv \frac{\omega}{\phi^2}
   \left( \phi_{,i} \phi_{,j} - \frac{1}{2} g_{ij} b \right) +
   \frac{1}{\phi} (\phi_{,i;j} - g_{ij} \Box \phi) + g_{ij} \Lambda -
   2\frac{\partial\Lambda}{\partial b} \phi_{,i} \phi_{,j} \; .
\label{Tphi}
\end{equation}
When $\phi=$const., $T^{\phi}_{ij}$ reduces to the vacuum energy-momentum
tensor $T^{\rm vac}_{ij} = \Lambda \, g_{ij}$, as expected.
The field equation for $\phi$ is
\begin{equation}
R + 2\Lambda + 2 \phi \frac{\partial\Lambda}{\partial\phi} -
   4\frac{\partial}{\partial x^l} \left( \phi
   \frac{\partial\Lambda}{\partial b} \right) \phi^{,l} - 4 \phi
   \frac{\partial\Lambda}{\partial b} \Box \phi =
   \frac{\omega}{\phi^2} b - \frac{2\omega}{\phi} \Box \phi -
   \frac{b}{\phi} \frac{d\omega}{d\phi} \; .
\label{FEphi}
\end{equation}
Eq.~(\ref{FEphi}) ensures that the conservation law $T^k_{i;k}=0$ holds,
as required.

When the scalar field $\phi$ depends only on $x^0 = ct$, then the
field equations for homogeneous and isotropic perfect-fluid cosmology
may be obtained from Eq.~(\ref{FEgij}) as follows.\cite{ref5,ref8}
The time-time component gives
\begin{equation}
\left( \frac{\dot{a}}{a} \right)^2 + \frac{k c^2}{a^2} -
   \frac{\Lambda c^2}{3} + \frac{2}{3}
   \frac{\partial\Lambda c^2}{\partial\dot{\phi}^2} \dot{\phi}^2 =
   \frac{8\pi}{3\phi c^2} \epsilon + \frac{\omega}{6} \left(
   \frac{\dot{\phi}}{\phi} \right)^2 - \frac{\dot{a}}{a}
   \frac{\dot{\phi}}{\phi} \; ,
\label{GST1}
\end{equation}
while the space-space components lead to
\begin{equation}
2 \frac{\ddot{a}}{a} + \left( \frac{\dot{a}}{a} \right)^2 + 
   \frac{k c^2}{a^2} - \Lambda c^2 = - \frac{8\pi}{\phi c^2} p -
   \frac{\omega}{2} \left( \frac{\dot{\phi}}{\phi} \right)^2 -
   \frac{\ddot{\phi}}{\phi} - 2\frac{\dot{a}}{a} 
   \frac{\dot{\phi}}{\phi} \; .
\label{GST2}
\end{equation}
From Eq.~(\ref{FEphi}), we have\footnote{This equation should be identical
with Eq.~(8) of Ref.~5.  The latter however contains two typesetting 
   mistakes, a missing $(\dot{a}/a)^2$-term and a missing overdot on 
   one of the $\phi$-terms.  Ref.~5 is also missing an equal sign between 
   the first and second lines of Eq.~(3), and the RHS of Eq.~(17) in that 
   paper should read $2\ddot{a}/\dot{a}$ rather than $2\ddot{a}/a$.}
\begin{eqnarray}
\frac{\ddot{a}}{a} + 
   \left( \frac{\dot{a}}{a} \right)^2 \!\! & \!\! + \!\! & \!\!
   \frac{k c^2}{a^2} - \frac{\Lambda c^2}{3} -
   \frac{\phi}{3} \frac{\partial\Lambda c^2}{\partial\phi} + 
   \frac{2}{3} \frac{\partial}{\partial t} \left( \phi
   \frac{\partial\Lambda c^2}{\partial\dot{\phi}^2} \right) \dot{\phi} + 
   \frac{2}{3} \phi \frac{\partial\Lambda c^2}{\partial\dot{\phi}^2}
   \left( \ddot{\phi} +3 \frac{\dot a}{a} \dot{\phi} \right) \nonumber \\
\!\! & \!\! = \!\! & \!\! \omega \frac{\dot{a}}{a} \frac{\dot{\phi}}{\phi} +
   \frac{\omega}{3} \frac{\ddot{\phi}}{\phi} +
   \frac{1}{6} \frac{d\omega}{d\phi} \frac{\dot{\phi}^2}{\phi} -
   \frac{\omega}{6} \left( \frac{\dot{\phi}}{\phi} \right)^2 \; .
\label{GST3}
\end{eqnarray}
This equation can be greatly simplified if the cross-terms in $a$ and
$\phi$ on either side are equated.  In particular the ansatz
\begin{equation}
\frac{\partial\Lambda c^2}{\partial\dot{\phi}^2} = \frac{\omega}{2\phi^2} \; ,
\label{ansatz}
\end{equation}
reduces Eq.~(\ref{GST3}) to the separable form
\begin{equation}
\frac{\ddot{a}}{a} + \left( \frac{\dot{a}}{a} \right)^2 +
   \frac{k c^2}{a^2} = \frac{1}{3} \left[ \Lambda c^2 + \phi
   \frac{\partial\Lambda c^2}{\partial\phi} - \frac{1}{2} 
   \frac{d\omega}{d\phi} \frac{\dot{\phi}^2}{\phi} + \frac{\omega}{2}
   \left( \frac{\dot{\phi}}{\phi} \right)^2 \right] \; .
\label{GSTsep}
\end{equation}
This is natural because one expects that $a$ (a metric quantity) and
$\phi$ (part of the source term $T^{\phi}_{ij}$) should be independent.
In what follows, we use Eqs.~(\ref{GST1}), (\ref{GST2}) and (\ref{GSTsep})
with the ansatz~(\ref{ansatz}) to solve for $a$, $\phi$ and $\Lambda c^2$.

\section{Connection with STM Theory and Evolution of the Scale Factor}
\noindent
Both sides of Eq.~(\ref{GSTsep}) may be equated to a parameter $K$ which
is taken to be constant according to the arguments of Ref.~8.  We focus
in this paper on the case $K=0$; the complementary situation in which
$K\neq0$ has been examined in Ref.~8.  The choice $K=0$ is motivated by
STM theory, a generalization of Kaluza-Klein theory in which Kaluza's
``cylinder condition'' is relaxed to allow dependence on the extra
coordinate(s) in principle.  A large literature\cite{ref6} has now built
up around this theory, in which 4D field equations of the form~(\ref{FEgij}),
together with a scalar field $\phi$ and a wide class of 4D matter fields,
can be ``induced'' from pure geometry in 5D; that is, from the action
principle $0 = \delta\int R^{\, (5)} \sqrt{-g^{\, (5)}} \, d^{\, 5}x$
(or, equivalently, the vacuum field equations $R^{\, (5)}_{AB}=0$,
where $A,B$ run over 0,1,2,3 and 5).

The energy-momentum tensor of the matter so induced is of a general nature
but has well-defined properties (see Refs.~6 for discussion) consistent with
those of the energy-momentum tensor $T^{\phi}_{ij}$ in Eq.~(\ref{Tphi}).
In this sense one can interpret both 4D matter and the scalar field in
modified GST theory as manifestations of 5D geometry.
More generally, one can consider homogeneous and isotropic extensions
of the Robertson-Walker metric to a 5D manifold,
$d^{\, (5)} s^2 = d^{\, (4)} s^2 + e^{\mu} (dx^{\, 5})^2$, where
$d^{\, (4)} s^2$ is the 4D RW line element and $\mu$ is taken as a
function of both $x^0=ct$ and the fifth coordinate $x^5$.\footnote{This
   new coordinate $x^5$ need not necessarily be lengthlike, as in
   traditional Kaluza-Klein theory, but could for instance be related
   to particle rest mass via $x^5=Gm/c^2$ (Ref.~10) or other quantities
   (see Refs.~5, 6 for discussion).}
As was first shown in Ref.~11 (see also Ref.~5), substitution of this 
metric into the 5D vacuum field equations leads to
\begin{equation}
\left( \frac{\dot{a}}{a} \right)^2 + \frac{kc^2}{a^2} + \frac{1}{2} 
   \frac{\dot{a}}{a} \dot{\mu} = 0 \; ,
\end{equation}
\begin{equation}
2 \frac{\ddot{a}}{a} + \left( \frac{\dot{a}}{a} \right)^2 + 
   \frac{kc^2}{a^2} + \frac{\ddot{\mu}}{2} + \frac{\dot{\mu}^2}{4} +
   \frac{\dot{a}}{a} \dot{\mu} = 0 \; ,
\end{equation}
\begin{equation}
\frac{\ddot{a}}{a} + \left( \frac{\dot{a}}{a} \right)^2 +
   \frac{kc^2}{a^2} = 0 \; .
\label{STM3}
\end{equation}
These results have the same form as the GST equations~(\ref{GST1}),
(\ref{GST2}) and (\ref{GSTsep}) respectively; and in fact all six
equations form a self-consistent set if the metric coefficient of the
fifth coordinate satisfies $\dot{\mu}=2\ddot{a}/\dot{a}$ (a result 
which played an important role in Ref.~5 but will not be needed here).
The salient point here is that Eq.~(\ref{STM3}) from 5D STM theory
gives the same result as the separable differential equation~(\ref{GSTsep})
in our modified GST theory, if the separability parameter $K$ in the latter
is set to zero.

The solution of Eq.~(\ref{STM3}) is\cite{ref11}
\begin{equation}
a = \sqrt{-kc^2 t^2 + \alpha t + \beta} \; ,
\label{scalefac}
\end{equation}
where $\alpha$ and $\beta$ are constants.  Differentiating twice, we find
that the scale factor, Hubble parameter $H\equiv\dot{a}/a$ and deceleration
parameter $q\equiv-a\ddot{a}/\dot{a}^2$ of the theory may be expressed in
dimensionless form as follows
\begin{equation}
\tilde{a}(\tilde{t}) = \sqrt{-k\tilde{t}^2 + \tilde{\alpha}\tilde{t} +
   \tilde{\beta}} \; , \; \; \;
\tilde{H}(\tilde{t}) = \frac{\tilde{\alpha} - 2k\tilde{t}}{2\tilde{a}^2} 
   \; , \; \; \;
q(\tilde{t}) = \frac{\tilde{\alpha}^2/4 + k\tilde{\beta}}
   {\tilde{H}^2\tilde{a}^4} \; ,
\end{equation}
where tildes denote dimensionless quantities and
\begin{equation}
\tilde{t} \equiv H_0t \; , \; \; \;
\tilde{a} \equiv \frac{H_0 a}{c} \; , \; \; \;
\tilde{H} \equiv \frac{H}{H_0} \; , \; \; \;
\tilde{\alpha} \equiv \frac{H_0\alpha}{c^2} \; , \; \; \;
\tilde{\beta} \equiv \frac{H_0^2\beta}{c^2} \; .
\end{equation}
We require that $\tilde{a}(\tilde{t}_0)=\tilde{a}_0$,
$\tilde{H}(\tilde{t}_0)=1$, and $q(\tilde{t}_0)=q_0$ at the present time
$\tilde{t}_0$, and also that the present phase of expansion begin in either
a big bang [$\tilde{a}(0)=0$] or a ``big bounce'' [$\tilde{H}(0)=0$] at
time zero.  The theory then admits four classes of solutions (Table~1).
\begin{table}[htbp]
\tcaption{Cosmological Model Parameters.}
\begin{tabular}{c c c c c c c}\\
\hline
Model & $k$ & $q_0$ & $\tilde{a}_0$ & $\tilde{\alpha}$ & $\tilde{\beta}$ &
   $\tilde{t}_0$ \\
\hline
I & $0$ & $1$ & free & $2\tilde{a}_0^2$ & 0 & $1/2$ \\
II & $+1$ & $q_0>1$ & $1/(\sqrt{q_0-1})$ & $2\sqrt{q_0} \, \tilde{a}_0^2$
   & $0$ & $1/(1+\sqrt{q_0})$ \\
III & $-1$ & $0<q_0<1$ & $1/(\sqrt{1-q_0})$ & $2\sqrt{q_0} \, \tilde{a}_0^2$
   & $0$ & $1/(1+\sqrt{q_0})$ \\
IV & $-1$ & $q_0<0$ & $1/(\sqrt{1-q_0})$ & $0$ & $-q_0/(1-q_0)^2$
   & $1/(1-q_0)$ \\
\hline\\
\end{tabular}
\end{table}
Models~I, II and III are all characterized by a big bang, as usual. 
For Model~IV, one also finds that the size of the scale factor at the 
moment of the big bounce is given by
$\tilde{a}(0)=\tilde{\beta}^{1/2}=\sqrt{-q_0} \, \tilde{a}_0^2$.
This tends to zero as $q_0 \rightarrow 0$, so that one also recovers 
a big bang in this limit.

Current data on Hubble's constant and the age of the Universe imply
that $H_0 \geq 65$~km~s$^{-1}$Mpc$^{-1}$ and $t_0 \geq 11$~Gyr,\cite{ref12}
so that we may take $\tilde{t}_0 > 0.7$.  This would rule out models of
types~I and II, which have $\tilde{t}_0 \leq 0.5$.  The evolution of the
scale factor in all four models is plotted in Fig.~1, where we have
adopted the values $q_o=2$ for Model~II, $\tilde{t}_0=0.8$ for Model~III,
and $q_0=-0.5$ for Model~IV.

The physical meaning of the deceleration parameter $q$ in scalar-tensor
gravity differs somewhat from that in standard general relativity,
and needs to be clarified before numerical boundary conditions can be
applied.  From the matter conservation law ${T^k}_{i;k}=0$ one obtains
the usual expression for the energy density of matter,
\begin{equation}
\epsilon = \epsilon_{\gamma} a^{-3(1+\gamma)} \; ,
\label{eps-mat}
\end{equation}
where $\gamma\equiv{p/\epsilon}$ and $\epsilon_\gamma=$~const.
The energy density and pressure of the scalar field can be obtained
from Eq.~(\ref{Tphi}) and read
\begin{equation}
\epsilon_{\phi} = \frac{\phi c^2}{8\pi}
   \left(-3 \frac{\dot{a}}{a} \frac{\dot{\phi}}{\phi} + 
   \frac{f_{\Lambda}}{\phi} \right) \; , \; \; \;
p_{\phi} = \frac{\phi c^2}{8\pi}
   \left( \frac{\ddot{\phi}}{\phi} + 2\frac{\dot{a}}{a}
   \frac{\dot{\phi}}{\phi} - \frac{f_{\Lambda}}{\phi} \right) \; .
\end{equation}
Eliminating $\phi$ from these two expressions, we find
\begin{equation}
\epsilon_{\phi}-3p_{\phi}=(3\gamma-1)\epsilon_{\gamma}a^{-3(\gamma+1)} \; .
\end{equation}
Following the approach adopted in other quintessence-type theories,\cite{ref4}
one can define a ratio of $p_{\phi}$ to $\epsilon_{\phi}$ for the scalar
field (analogous to that for matter) via
\begin{equation}
p_{\phi}\equiv \gamma_{\phi}\epsilon_{\phi} \; ,
\label{phiEOS}
\end{equation}
where $\gamma_{\phi}$ is in general a function of $t$. More sophisticated 
equations of state are possible too. With Eq.~(\ref{phiEOS}), one finds
for the deceleration parameter,
\begin{equation}
q = \frac{(1+3\gamma)}{2} \Omega_m + \frac{(1+3\gamma_{\phi})}{2}
   \Omega_{\phi} \; ,
\end{equation}
where the energy density parameters of matter and the scalar field 
are respectively
\begin{equation}
\Omega_m \equiv
   \frac{8\pi\epsilon}{3H^2\phi c^2} \; , \; \; \;
\Omega_{\phi} \equiv
   \frac{8\pi\epsilon_{\phi}}{3H^2\phi c^2} = -\frac{1}{H}
   \frac{\dot{\phi}}{\phi} + \frac{f_{\Lambda}}{3H^2\phi} \; .
\label{Omega-mat}
\end{equation}
In the case $\phi=$~const., [or equivalently $\Lambda=$const.] where 
the energy-momentum tensor of the scalar field reduces to that of a vacuum
($\gamma_{\phi}=\gamma_{\rm vac}=-1$) and we recover the standard expression
for the deceleration parameter,
\begin{equation}
q = \frac{(1+3\gamma)}{2} \Omega_m - \Omega_{\rm vac} \; .
\label{decel}
\end{equation}
Observations of galaxy clusters indicate that $\Omega_{m,0} < 0.5$, while
analysis of fluctuations in the cosmic microwave background (CMB) gives
$\Omega_{m,0}+\Omega_{\Lambda,0} \approx 1.1 \pm 0.1$ (see Ref.~12 for a
recent review of the observational data).  Both numbers together imply that
$\Omega_{\Lambda,0} \geq 0.5$, a result which agrees with data on distant
supernovae.  For dustlike ($\gamma=0$) models with $\phi=$~const., we would
then conclude from Eq.~(\ref{decel}) that $q_0 < -0.2$, which would restrict
us to Model~IV in our theory.  For more general situations, we can say
nothing about the value of $q_0$ at this stage, and either of Models~ III
and IV are viable in principle.

\section{Evolution of the Scalar Field}
\noindent
Integrating Eq.~(\ref{ansatz}) with $\omega=\omega(\phi)$, we find with
the help of Eq.~(\ref{STM3}) that
\begin{equation}
\Lambda c^2 = \frac{\omega}{2} \left( \frac{\dot{\phi}}{\phi} \right)^2 +
   \frac{f_{\Lambda}}{\phi} \; ,
\label{Lambda}
\end{equation}
where $f_{\Lambda}=$~const.  Eq.~(\ref{Lambda}) shows that in an RW 
universe with $\phi=$~const., $\Lambda$ is constant too.
Alternatively, in Brans-Dicke (BD) theory with $\phi$ a power-law
function of time and $\omega=$const., the first term on the right-hand
side of Eq.~(\ref{Lambda}) goes as $t^{-2}$ (this behavior is common
to other modified GST theories\cite{ref7}).

The scalar field $\phi(ct)$ is then obtained from either of
Eqs.~(\ref{GST1}) or (\ref{GST2}), with the help of Eq.~(\ref{ansatz})
for $\partial\Lambda c^2/\partial\dot{\phi}^2$, Eq.~(\ref{scalefac})
for $a$, Eq.~(\ref{eps-mat}) for $\epsilon$ (or $p$), and
Eq.~(\ref{Lambda}) for $\Lambda c^2$.  The result is
\begin{equation}
\dot{\phi} + \frac{\alpha^2+4k\beta c^2}{2(-2kc^2t+\alpha)a^2} \phi =
   \frac{2a^2}{-2kc^2t+\alpha} \left[ \frac{8\pi\epsilon_{\gamma}}{3c^2}
   a^{-3(1+\gamma)} + \frac{f_{\Lambda}}{3} \right] \; .
\label{phiDE}
\end{equation}
Eq.~(\ref{phiDE}) is a linear differential equation of first order,
and may be solved analytically for $\phi(t)$ in the cases $k=0,\pm1$ and
$\gamma=-1, 1, 1/3,0$.  For $k=0$ we obtain
\begin{equation}
\phi = \left\{ \begin{array}{l}
   \frac{\textstyle 32\pi\epsilon_{-1}}{\textstyle 15\alpha^2 c^2} a^4+
      \frac{\textstyle 4f_{\Lambda}}{\textstyle 15\alpha^2} a^4+
      \frac{\textstyle \phi_k}{\textstyle a}
      \hspace{5mm} \mbox{if $\gamma=-1$} \\ \\
  -\frac{\textstyle 32\pi\epsilon_1}{\textstyle 3\alpha^2 c^2 a^2}+
      \frac{\textstyle 4f_{\Lambda}}{\textstyle 15\alpha^2} a^4+
      \frac{\textstyle \phi_k}{\textstyle a}
      \hspace{5mm} \mbox{if $\gamma=1$} \\ \\
  \frac{\textstyle 32\pi\epsilon_{1/3}}{\textstyle 3\alpha^2 c^2}+
      \frac{\textstyle 4f_{\Lambda}}{\textstyle 15\alpha^2} a^4+
      \frac{\textstyle \phi_k}{\textstyle a}
      \hspace{5mm} \mbox{if $\gamma=\frac{1}{3}$} \\ \\
  \frac{\textstyle 16\pi\epsilon_0 t}{\textstyle 3\alpha c^2 a}+
      \frac{\textstyle 4f_{\Lambda}}{\textstyle 15\alpha^2} a^4+
      \frac{\textstyle \phi_k}{\textstyle a}
      \hspace{5mm} \mbox{if $\gamma=0$} \; . \end{array} \right.
\label{phi0}
\end{equation}
We solve Eq.~(\ref{phiDE}) for $k=\pm1$ as follows,
\begin{equation}
\phi=\left\{ \begin{array}{l}
   \frac{\textstyle \mp 2c^2t+\alpha}{\textstyle a} 
       \left( \frac{\textstyle 16\pi\epsilon_{-1}}{\textstyle 3c^2}
       \Phi + \frac{\textstyle 2f_{\Lambda}}{\textstyle 3} \Phi +
       \phi_k \right)
       \hspace{5mm} \mbox{if $\gamma=-1$} \\ \\
   \frac{\textstyle 32\pi\epsilon_1 (-8c^4t^2\pm 8\alpha c^2t-
       \alpha^2\pm 4\beta c^2)}{\textstyle 3c^2(\alpha^2\pm 4\beta c^2)^2a^2}+
       \frac{\textstyle \mp 2c^2t+\alpha}{\textstyle a}
       \left( \frac{\textstyle 2f_{\Lambda}}{\textstyle 3} \Phi +
       \phi_k \right)
       \hspace{5mm} \mbox{if $\gamma=1$} \\ \\
   \frac{\textstyle 32\pi\epsilon_{1/3}}{\textstyle 3c^2(\alpha^2\pm 4
       \beta c^2)} + \frac{\textstyle \mp2c^2t+\alpha}{\textstyle a}
       \left( \frac{\textstyle 2f_{\Lambda}}{\textstyle 3} \Phi +
       \phi_k \right)
       \hspace{5mm} \mbox{if $\gamma=\frac{1}{3}$} \\ \\
   \pm\frac{\textstyle 8\pi\epsilon_0}{\textstyle 3c^4a} +
       \frac{\textstyle \mp2c^2t+\alpha}{\textstyle a}
       \left( \frac{\textstyle 2f_{\Lambda}}{\textstyle 3} \Phi +
       \phi_k \right)
       \hspace{5mm} \mbox{if $\gamma=0$ ,} \end{array} \right.
\label{phi1}
\end{equation}
where
\begin{equation}
\Phi \equiv \frac{a^3}{-4c^4 t \pm 2\alpha c^2} + 
   \frac{3}{16c^4} \left[ (\mp 2c^2 t + \alpha) a - 
   \frac{\alpha^2 \pm 4\beta c^2}{2} I_F \right] \; ,
\end{equation}
and
\begin{equation}
I_F \equiv \left\{ \begin{array}{l}
   -\frac{\textstyle 1}{\textstyle c} \arcsin
      \frac{\textstyle -2c^2t+\alpha}{\textstyle \sqrt{\alpha^2+4\beta c^2}}
      \hspace{5mm} \mbox{if $k=+1$} \\ \\
   \frac{\textstyle 1}{\textstyle 2c} \ln \left|
      \frac{\textstyle 2c^2t+\alpha+2ca}{\textstyle 2c^2t+\alpha-2ca} \right|
      \hspace{5mm} \mbox{if $k=-1$} \; . \end{array} \right.
\end{equation}
To obtain a quantitative idea of the evolution of the scalar field with
time, we need the parameter $\epsilon_{\gamma}$ in Eqs.~(\ref{phi0}) and
(\ref{phi1}).  Application of Eq.~(\ref{eps-mat}) at the present time with
the boundary condition
$\epsilon(t_0)= c^2 \rho_0 = c^2 \rho_{\rm crit} \Omega_{m,0}$ leads to
\begin{eqnarray}
\epsilon_{-1} = \left( \frac{3H_0^2c^2}{8\pi G_0} \right) \Omega_{m,0}
   \; \!\! & \!\! , \!\! & \!\! \; \; \;
\epsilon_{1} = \left( \frac{3c^8}{8\pi G_0H_0^4} \right)
   \Omega_{m,0} \tilde{a}_0^6 \; , \nonumber \\
\epsilon_{1/3} = \left( \frac{3c^6}{8\pi G_0H_0^2} \right) \Omega_{m,0}
   \tilde{a}_0^4 \; \!\! & \!\! , \!\! & \!\! \; \; \;
\epsilon_{0} = \left( \frac{3c^5}{8\pi G_0H_0} \right)
   \Omega_{m,0} \tilde{a}_0^3 \; ,
\label{epsBCs} 
\end{eqnarray}
where the present critical density $\rho_{\rm crit}\equiv 3H_0^2/8\pi G_0$
and $G_0$ is the present value of Newton's gravitational constant
(which varies roughly as $1/\phi$ in scalar-tensor theories).
Of these parameters, only the last is relevant to the present (dustlike)
universe.  In a more sophisticated model one would allow for different
epochs (characterized by different values of $\gamma$) and evaluate
Eq.~(\ref{eps-mat}) across the phase transitions between them, enforcing
continuity to obtain realistic values of $\epsilon_{\gamma}$.

We proceed to define a dimensionless scalar field via
$\tilde{\phi} \equiv (G_0/\Omega_{m,0})\phi$.  With $\epsilon_{\gamma}$
as given in Eqs.~(\ref{epsBCs}), this is found to consist of three
components,
\begin{equation}
\tilde{\phi} = \tilde{\phi}_1 + \tilde{\phi}_2 + \tilde{\phi}_3 \; ,
\label{3terms}
\end{equation}
where $\tilde{\phi}_1$, $\tilde{\phi}_2$ and $\tilde{\phi}_3$ take simple
analytic forms, depending on the model parameters (see Appendix).
Figs.~2(a)--(d) show the evolution of the scalar field $\tilde{\phi}$
in Models I--IV respectively, using the same values for $q_0$ as those
in Fig.~1, and values of $\tilde{f}_{\Lambda}$ and $\tilde{\phi}_k$
as marked beside the curves.

\section{Evolution of the Cosmological Term}
\noindent
The cosmological term is found from Eq.~(\ref{Lambda}) and may be expressed
in the usual form of a dimensionless vacuum energy density parameter
\begin{equation}
\Omega_{\Lambda} \equiv \frac{\Lambda c^2}{3H_0^2} = \frac{\omega}{6} 
   \left( \frac{1}{H_0} \frac{\dot{\tilde{\phi}}}{\tilde{\phi}}
   \right)^2 + \frac{5\tilde{f}_{\Lambda}}{\tilde{\phi}} \; .
\label{OmegaLam}
\end{equation}
(Note that this is  not the same as $\Omega_{\phi}$.)
At the present time, $\Omega_{\Lambda}=\Omega_{\Lambda,0}$ is the quantity
measured by observers using, e.g., the magnitude-redshift relation for
distant supernovae.  Eq.~(\ref{OmegaLam}) cannot be evaluated without
a specific functional form for $\omega(\phi)$.  Let us consider to begin
with the case in which $\omega=\omega_0=$~const., and the simplest models
in which $\tilde{f}_{\Lambda} = \tilde{\phi}_k=0$.  Taking $\gamma=0$
(dust) and $k=0$ (Model~I), we find from Eq.~(\ref{OmegaLam}) that
\begin{equation}
\Omega_{\Lambda,0} = \frac{\omega_0}{6} \left(
   \frac{1-\tilde{t}_0}{\tilde{t}_0} \right)^2 = \frac{\omega_0}{6} \; ,
\end{equation}
since $\tilde{t}_0=1/2$ for Model~I.  Using the observational upper limit
$\Omega_{\Lambda,0} \leq 1.1$\cite{ref12} we find that $\omega_0 < 7$,
which violates the current experimental lower bound ($\omega_0 \geq 600$)
on this quantity.\cite{ref13}  This means that, at least in the case
$k=0$ with $\omega=$~const., we cannot have both $\tilde{f}_{\Lambda}$
and $\tilde{\phi}_k$ equal to zero.

Alternatively, we can use Eq.~(\ref{OmegaLam}) to constrain the values
of $\tilde{f}_{\Lambda}$ and $\tilde{\phi}_k$ if we impose the boundary
conditions $\Omega_{\Lambda}=\Omega_{\Lambda,0}$ and $\omega=\omega_0=600$
(say).  For Model~I ($k=0$) as the simplest case, we find on 
differentiating Eq.~(\ref{3terms}) that this procedure gives a quadratic 
for $\tilde{f}_{\Lambda}$ (if $\tilde{\phi}_k$ is held constant) or
$\tilde{\phi}_k$ (if $\tilde{f}_{\Lambda}$ is held constant).
In particular, if $\tilde{f}_{\Lambda}=0$, then $\tilde{\phi}_k=0.418$
and $0.598$ both give $\Omega_{\Lambda,0}=0.8$ at the present time
(with minima in the past and future respectively).  Similarly, if
$\tilde{\phi}_k=0$, then the same thing occurs for
$\tilde{f}_{\Lambda}=-0.140$ and $-0.110$.  This second possibility
however involves a negative cosmological term in the near past (or future),
which is strongly disfavored by observation as it would shorten the age
of the Universe.  The implication is that $\tilde{\phi}_k$, which arises
as a constant of integration in Eqs.~(\ref{phi0}) and (\ref{phi1}),
is probably nonzero in models of this kind if they are to be realistic.
This scalar-field theory, in other words, is of the ``chaotic'' kind,
in which the field begins to roll from an (arbitrary) nonzero initial
value.\cite{ref14}

We show four representative solutions of Eq.~(\ref{OmegaLam}) in Fig.~3,
which assumes Model~I ($k=0$) and $\omega_0=600$.  These have been fit to
$\Omega_{\Lambda,0}=0.8$ at the present time, as suggested by current
experimental data.\cite{ref12}  Simplest is the $\tilde{f}_{\Lambda}=0$
solution, but all are natural in the sense of having dimensionless
parameters (other than $\omega_0$) of order unity.  All models, moreover,
show a steep increase in $\Omega_{\Lambda}$ in the past direction.
This behavior is generic to constant-$\omega$ solutions of Model~II, III
and IV as well as those of Model~I.  The modified GST theory, therefore,
provides one possible mechanism for addressing the problem of the
cosmological ``constant''.\cite{ref3}  Any such increase during the
nucleosynthesis and decoupling eras, however, must satisfy constraints
based on light element synthesis\cite{ref7,ref15} and the observed
spectrum of CMB anisotropies.\cite{ref16}  In a realistic GST, the
behavior of $\Omega_{\Lambda}$ is probably more complicated than that
shown in Fig.~3, and $\omega(\phi)$ will almost certainly {\em not\/}
be constant.

Let us accordingly consider some aspects of models in which
$\omega(\phi) \neq$~const.  We focus on the matter and radiation eras,
since these may be constrained by observational cosmology,\cite{ref4}
and again take the simplest case with $k=0$ and $\beta=0$.  During the
matter era ($\gamma=0$), the cosmological term is given by
Eq.~(\ref{Lambda}) as
\begin{equation}
\Lambda_m c^2 = \frac{1}{8t_m^2} \left[ \omega(\phi_m) \left( \frac{
   1 + 4E_1 t_m^{3/2} - E_2/t_m}{1 + E t_m^{3/2} + E_2/t_m} \right)^2 + 
   \frac{30E_1 t_m^{3/2}}{1 + E_1 t_m^{3/2} + E_2/t_m} \right] \; ,
\label{Lambda-mat}
\end{equation}
where
\begin{eqnarray}
E_1 \equiv \frac{f_{\Lambda} \alpha^{3/2} c^2}{20\pi\epsilon_0} \; , \; \; \;
E_2 \equiv \frac{3\phi_k\alpha c^2}{16\pi\epsilon_0} \; .
\label{E1E2}
\end{eqnarray}
Here $\epsilon_0$ is the value of $\epsilon_\gamma$ at $\gamma=0$ in 
Eq.~(\ref{eps-mat}) [see Eq.~(\ref{epsBCs})].  The energy density parameter
of matter may be obtained from Eq.~(\ref{Omega-mat})
\begin{equation}
\Omega_m = \frac{2}{1 + E_1 t_m^{3/2} + E_2/t_m} \; .
\end{equation}
Its present value, $\Omega_{m,0} = 2/(1 + E_1 t_0^{3/2} + E_2/t_0)$,
is constrained experimentally to the range
$0.01 \leq \Omega_{m,0} \leq 0.5$.\cite{ref12}

During the radiation era ($\gamma=1/3$), on the other hand,
\begin{equation}
\Lambda_r c^2 = \frac{1}{4t_r^2} \left[ 8\omega(\phi_r) \left( \frac{ B_1
   t_r^2 - B_2/4 t_r^{1/2}}{1 + B_1 t_r^2 + B_2/t_r^{1/2}} \right)^2 +
   \frac{15B_1 t_r^2}{1 + B_1 t_r^2 + B_2/t_r^{1/2}} \right] \; ,
\label{Lambda-rad}
\end{equation}
where
\begin{equation}
B_1 t_r^2 \equiv \frac{f_{\Lambda} \alpha^2 c^2}{40\pi\epsilon_{1/3}} t_r^2 =
   \frac{\epsilon_m}{2\epsilon_r} E_1 t_m^{3/2} \; , \; \; \;
\frac{B_2}{t_r^{1/2}} \equiv \frac{3\phi_k \alpha^{3/2} c^2}{32\pi
   \epsilon_{1/3} t_r^{1/2}} = \frac{\epsilon_m}{2\epsilon_r}
   \left( \frac{t_m}{t_r} \right)^{5/2} \frac{E_2}{t_m} \; .
\label{B1B2}
\end{equation}
If we take $f_\Lambda=0$ as above (i.e., $E_1=0$ and $B_1=0$), then the
cosmological term at nucleosynthesis time $t=t_n$ is given by
Eq.~(\ref{Lambda-rad}) as
\begin{eqnarray}
\Lambda_n c^2 = \frac{\omega(\phi_n)}{8t_n^2}
   \left( \frac{B_2/t_n^{1/2}}{1 + B_2/t_n^{1/2}} \right)^2
   \sim \frac{\omega(\phi_n)}{8t_n^2} \; ,
\end{eqnarray}
where  $B_2/t_n^{1/2} \gg 1$ from Eqs.~(\ref{E1E2}) and (\ref{B1B2}), since 
$\epsilon_n \sim 10^{30}\epsilon_0$, $t_0\sim 10^{15}t_n$ and $E_2/t_0\sim 4$.
At decoupling time $t=t_d$, finally, Eq.~(\ref{Lambda-mat}) gives
\begin{eqnarray}
\Lambda_d c^2 = \frac{\omega(\phi_d)}{8t_d^2}
   \left( \frac{1-E_2/t_d}{1 + E_2/t_d} \right)^2
   \sim \frac{\omega(\phi_d)}{8t_d^2} \; ,
\end{eqnarray}
where $E_2/t_d = (2/\Omega_{m,0}-1)(t_0/t_d) \gg 1$ from Eq.~(\ref{E1E2}),
since $t_0\sim 10^5t_d$.  These expressions can in principle be used to 
constrain the behavior of $\omega(\phi)$ during the radiation era via
nucleosynthesis considerations\cite{ref7,ref15} and observations of the
power spectrum of CMB fluctuations.\cite{ref16}

\section{Discussion}
\noindent
We have modified the generalized scalar-tensor theory by taking the
cosmological term as a function not only of the scalar field but of its
gradient as well.  This means that we effectively interpret both the
second and third terms in Wagoner's Lagrangian density [Eq.~(2) in Ref.~1]
as a single variable cosmological term.  In the context of higher-dimensional
theories this may be justified by considering the origin of the scalar
field, which is induced in 4D spacetime by the geometry of the empty
higher-dimensional universe.\cite{ref17}  The scalar field and cosmological
term can in this sense be recognized together as a single geometrical entity.
We in the 4D universe then study this as constant for de~Sitter models, or
as temperature-dependent in spontaneous symmetry-breaking models,\cite{ref18}
or as kinetic energy-like in connection with the age\cite{ref19} or dark
matter problems.\cite{ref20}  A recent comprehensive discussion of the
cosmological term has been given by Sahni and Starobinsky.\cite{ref21}

We have previously explored the relationship between the modified GST
theory~(\ref{GSTaction}) and 5D STM theory in Ref.~5.
There, an identification of the extra part of the higher-dimensional
metric tensor in terms of the scalar field $\phi$ enabled us to obtain
solutions for $\Lambda$ relevant to vacuum-dominated or stiff matter-like
phases in the history of the Universe.  In the present paper, a different
ansatz~(\ref{ansatz}) has led to a wider class of solutions which can be
used to describe the radiation and matter-dominated eras.  As seen from
Eq.~(\ref{Lambda}) and the solutions~(\ref{phi0}) and (\ref{phi1}),
the cosmological term evolves differently in each era because $\phi$
depends explicitly on $\gamma$.  This presents a mechanism for generating
new cosmological terms during successive cosmological phase transitions.

This work has made use of STM theory, because it is one of our motivations
to explore the relevance of higher-dimensional physics to 4D cosmology.  
But Eq.~(\ref{GSTsep}) admits a second class of solutions for the
evolution of the scale factor when we do not make use of Eq.~(\ref{STM3})
from STM theory to set Eq.~(\ref{GSTsep}) to zero.  Fukui et al.\cite{ref8}
have used these to model inflation in the early universe --- something for
which the present set of solutions is not suited, as seen from the form
of Eq.~(\ref{scalefac}).  Taken together, both classes of solutions can
allow us to follow the evolution of the cosmological term throughout the
history of the Universe.  These and other aspects of modified GST theory
will be pursued further elsewhere.

\nonumsection{Acknowledgements}
\noindent
T.~F. is grateful for the hospitality of the Department of Physics,
University of Waterloo, while staying on the research program of
Dokkyo University.  J.~M.~O. is supported by a Postdoctoral Fellowship
from the Japan Society for the Promotion of Science (JSPS).

\nonumsection{References}

\newpage
\appendix

\noindent
In this Appendix we give expressions for the components $\tilde{\phi}_1$,
$\tilde{\phi}_2$ and $\tilde{\phi}_3$ of the dimensionless scalar field,
Eq.~(\ref{3terms}), as obtained from Eqs.~(\ref{phi0}) and (\ref{phi1})
and plotted in Figs.~2 of the main text.  For Model I:
\begin{eqnarray}
\tilde{\phi}_1 \!\! & \!\! = \!\! & \!\! \left\{ \begin{array}{l}
   H_{0}t\left(\frac{\textstyle a}{\textstyle a_0} \right)^{-1}
      \hspace{5mm} \mbox{ if $\gamma=0$} \\ \\
   1 \hspace{5mm} \mbox{ if $\gamma=1/3$} \\ \\
      -\left( \frac{\textstyle a}{\textstyle a_0} \right)^{-2}
      \hspace{5mm} \mbox{ if $\gamma=1$} \\ \\
   \frac{1}{5} \left( \frac{\textstyle a}{\textstyle a_0} \right)^4
      \hspace{5mm} \mbox{ if $\gamma=-1$ ;} \end{array} \right. 
      \nonumber \\ \nonumber \\
\tilde{\phi}_2 \!\! & \!\! = \!\! & \!\! 5\tilde{f}_{\Lambda}
   \tilde{\phi}_1 \mbox{\hspace{5mm}} \mbox{(with } \tilde{\phi}_1
   \mbox{ as for }\gamma=-1 \mbox{ case)} \; ;
   \nonumber \\ \nonumber \\
\tilde{\phi}_3 \!\! & \!\! = \!\! & \!\! \tilde{\phi}_k
   \left( \frac{\textstyle a}{\textstyle a_0} \right)^{-1} \; .
\end{eqnarray}
For Model II:
\begin{eqnarray}
\tilde{\phi}_1 \!\! & \!\! = \!\! & \!\! \left\{ \begin{array}{l}
   \frac{\textstyle 1}{\textstyle q_0-1}
      \left( \frac{\textstyle a}{\textstyle a_0} \right)^{-1}
      \hspace{5mm} \mbox{ if $\gamma=0$} \\ \\
   \frac{\textstyle 1}{\textstyle q_0}
      \hspace{5mm} \mbox{ if $\gamma=1/3$} \\ \\
   \frac{\textstyle 1}{\textstyle q_0}
      \left( \frac{\textstyle a}{\textstyle a_0} \right)^{-2}
      \left( 1 - 2 \xi^2 \right)
      \hspace{5mm} \mbox{ if $\gamma=1$} \\ \\
   \frac{\textstyle 1}{\textstyle q_0-1}
      \left( \frac{\textstyle a}{\textstyle a_0} \right)^2
      \left( 1 +\frac{3}{2}\zeta^2 +
      \frac{\textstyle 3\,\zeta^3}{\textstyle 2\,\xi^2} \arcsin\xi \right)
      \hspace{5mm} \mbox{ if $\gamma=-1$ ;} \end{array} \right. 
      \nonumber \\ \nonumber \\
\tilde{\phi}_2 \!\! & \!\! = \!\! & \!\! 5\tilde{f}_{\Lambda} \tilde{\phi}_1
   \mbox{\hspace{5mm}} \mbox{(with } \tilde{\phi}_1
   \mbox{ as for }\gamma=-1 \mbox{ case)} \; ;
   \nonumber \\ \nonumber \\
\tilde{\phi}_3 \!\! & \!\! = \!\! & \!\! 2 \tilde{\phi}_k \zeta \; ,
\end{eqnarray}
For Model III:
\begin{eqnarray}
\tilde{\phi}_1 \!\! & \!\! = \!\! & \!\! \left\{ \begin{array}{l}
   \frac{\textstyle 1}{\textstyle q_0-1}
      \left( \frac{\textstyle a}{\textstyle a_0} \right)^{-1}
      \hspace{5mm} \mbox{ if $\gamma=0$} \\ \\
   \frac{\textstyle 1}{\textstyle q_0}
      \hspace{5mm} \mbox{ if $\gamma=1/3$} \\ \\
   \frac{\textstyle 1}{\textstyle q_0}
      \left( \frac{\textstyle a}{\textstyle a_0} \right)^{-2}
      \left( 1 - 2 \xi^2 \right) \hspace{5mm} \mbox{ if $\gamma=1$} \\ \\
   \frac{\textstyle 1}{\textstyle q_0-1}
      \left( \frac{\textstyle a}{\textstyle a_0} \right)^2
      \left( 1 -\frac{3}{2}\zeta^2 +
      \frac{\textstyle 3\,\zeta^3}{\textstyle 4\,\xi^2} \ln
      \left| \frac{\textstyle \zeta+1}{\textstyle \zeta-1} \right| \right)
      \hspace{5mm} \mbox{ if $\gamma=-1$ ;} \end{array} \right. 
      \nonumber \\ \nonumber \\
\tilde{\phi}_2 \!\! & \!\! = \!\! & \!\! 5\tilde{f}_{\Lambda} \tilde{\phi}_1
   \mbox{\hspace{5mm}} \mbox{(with } \tilde{\phi}_1
   \mbox{ as for }\gamma=-1 \mbox{ case)} \; ;
   \nonumber \\ \nonumber \\
\tilde{\phi}_3 \!\! & \!\! = \!\! & \!\! 2 \tilde{\phi}_k \zeta \; .
\end{eqnarray}
For Model IV:
\begin{eqnarray}
\tilde{\phi}_1 \!\! & \!\! = \!\! & \!\! \left\{ \begin{array}{l}
   \frac{\textstyle 1}{\textstyle q_0-1}
      \left( \frac{\textstyle a}{\textstyle a_0} \right)^{-1}
      \hspace{5mm} \mbox{ if $\gamma=0$} \\ \\
   \frac{\textstyle 1}{\textstyle q_0}
      \hspace{5mm} \mbox{ if $\gamma=1/3$} \\ \\
   \frac{\textstyle 1}{\textstyle q_0}
      \left( \frac{\textstyle a}{\textstyle a_0} \right)^{-2}
      \left( 1 + 2 \eta^2 \right)
      \hspace{5mm} \mbox{ if $\gamma=1$} \\ \\
   \frac{\textstyle 1}{\textstyle q_0-1}
      \left( \frac{\textstyle a}{\textstyle a_0} \right)^2
      \left( 1 -\frac{3}{2}\lambda^2 -
      \frac{\textstyle 3\,\lambda^3}{\textstyle 4\,\eta^2} \ln
      \left| \frac{\textstyle \lambda+1}{\textstyle \lambda-1} \right| \right)
      \hspace{5mm} \mbox{ if $\gamma=-1$ ;} \end{array} \right. \nonumber \\
      \nonumber \\
\tilde{\phi}_2 \!\! & \!\! = \!\! & \!\! 5\tilde{f}_{\Lambda} \tilde{\phi}_1
   \mbox{\hspace{5mm}} \mbox{(with } \tilde{\phi}_1
   \mbox{ as for }\gamma=-1 \mbox{ case)} \; ;
   \nonumber \\ \nonumber \\
\tilde{\phi}_3 \!\! & \!\! = \!\! & \!\! 2 \tilde{\phi}_k \lambda \; .
\end{eqnarray}
In these equations we have introduced four new parameters
\begin{eqnarray}
\xi \!\! & \!\! \equiv \!\! & \!\! 1-\left( 1 -
   \frac{\textstyle 1}{\textstyle \sqrt{q_0}} \right)
   \frac{\textstyle t}{\textstyle t_0} \; , \; \; \;
\zeta \equiv \xi \sqrt{\frac{\textstyle q_0}{\textstyle |q_0-1|}}
   \left( \frac{\textstyle a}{\textstyle a_0} \right)^{-1} \; , \nonumber \\
\eta \!\! & \!\! \equiv \!\! & \!\! \left( 
   \frac{\textstyle 1}{\textstyle \sqrt{-q_0}} \right)
   \frac{\textstyle t}{\textstyle t_0} \; , \; \; \;
\lambda \equiv {\frac{\textstyle 1}{\textstyle \sqrt{1-q_0}}}
   \left( \frac{\textstyle a}{\textstyle a_0} \right)^{-1}
   \frac{\textstyle t}{\textstyle t_0} \; ,
\end{eqnarray}
and additionally defined two dimensionless free parameters
$\tilde{f}_{\Lambda} \equiv (G_0/15 H_0^2\Omega_{m,0}) f_{\Lambda}$ and
$\tilde{\phi}_k \equiv (G_0/a_0 \Omega_{m,0}) \phi_k$ (if $k=0$) or
$\tilde{\phi}_k \equiv (cG_0/\Omega_{m,0}) \phi_k$ (if $k=\pm1$).
In a natural theory one expects both of these parameters to take values
of order unity (no fine-tuning).  As discussed in Section~5 of the main
text, this expectation appears to be borne out for the simplest models.

\end{document}